\documentclass[conference]{IEEEtran}
\IEEEoverridecommandlockouts
\usepackage{cite}
\usepackage{amsmath,amssymb,amsfonts}
\usepackage{algorithmic}
\usepackage{graphicx}
\usepackage{textcomp}
\usepackage{xcolor}
\usepackage{etoolbox}
\makeatletter
\patchcmd{\@makecaption}
  {\scshape}
  {}
  {}
  {}
\makeatletter

\def\BibTeX{{\rm B\kern-.05em{\sc i\kern-.025em b}\kern-.08em
    T\kern-.1667em\lower.7ex\hbox{E}\kern-.125emX}}

\usepackage{soul}
\usepackage{glossaries-extra}
\setabbreviationstyle[acronym]{long-short}
\newacronym{iot}{IoT}{Internet of Things}
\newacronym{plkg}{PLKG}{Physical Layer Key Generation}
\newacronym{aes}{AES}{Advanced Encryption Standard}
\newacronym{dh}{DH}{Diffie-Hellman}
\newacronym{rsa}{RSA}{Rivest-Shamir-Adleman}
\newacronym{qkd}{QKD}{Quantum Key Distribution}
\newacronym{pqc}{PQC}{Post Quantum Cryptography}
\newacronym{qos}{QoS}{Quality of Service}
\newacronym{pls}{PLS}{Physical Layer Security}
\newacronym{evtol}{eVTOL}{electric vertical take-off and landing}
\newacronym{csi}{CSI}{channel state information}
\newacronym{ue}{UE}{User Equipment}
\newacronym{uwb}{UWB}{Ultra Wideband}
\newacronym{lte}{LTE}{Long-Term Evolution}
\newacronym{tdd}{TDD}{Time Division Duplex}
\newacronym{dmrs}{DMRS}{Demodulation Reference Signals}
\newacronym{nic}{NICs}{Network Interface controllers}
\newacronym{sdr}{SDR}{Software-defined radio}
\newacronym{qpsk}{QPSK}{Quadrature Phase Shift Keying}

\makeatletter
\newcommand*{\rom}[1]{\expandafter\@slowromancap\romannumeral #1@}
\makeatother

\usepackage{tikz}
\usepackage{textcomp}
\usepackage{hyperref}
\usepackage{lipsum}

\newcommand\copyrighttext{%
  \footnotesize \textcopyright 2023 IEEE.  Personal use of this material is permitted.  Permission from IEEE must be obtained for all other uses, in any current or future media, including reprinting/republishing this material for advertising or promotional purposes, creating new collective works, for resale or redistribution to servers or lists, or reuse of any copyrighted component of this work in other works.
  }
\newcommand\copyrightnotice{%
\begin{tikzpicture}[remember picture,overlay]
\node[anchor=south,yshift=10pt] at (current page.south) {\fbox{\parbox{\dimexpr\textwidth-\fboxsep-\fboxrule\relax}{\copyrighttext}}};
\end{tikzpicture}%
}
    
\begin{document}

\title{Physical Layer Security in a Private 5G Network for Industrial and Mobility Application
}

\author{\IEEEauthorblockN{Shivraj Hanumant Gonde}
\IEEEauthorblockA{\textit{Secure Land Communications} \\
\textit{Airbus} \\
Ulm, Germany \\
shivraj\_hanumant.gonde@airbus.com} \\
\IEEEauthorblockN{Martin Kubisch}
\IEEEauthorblockA{\textit{Central Research and Technology} \\
\textit{Airbus}\\
Munich, Germany \\
martin.kubisch@airbus.com}
\and
\IEEEauthorblockN{Christoph Frisch}
\IEEEauthorblockA{\textit{Chair for Security in Information Technology} \\
\textit{Technical University of Munich}\\
Munich, Germany \\
chris.frisch@tum.de} \\
\IEEEauthorblockN{Thomas Meyerhoff}
\IEEEauthorblockA{\textit{Central Research and Technology} \\
\textit{Airbus}\\
Hamburg, Germany \\
thomas.meyerhoff@airbus.com}
\and
\IEEEauthorblockN{Svetoslav Duhovnikov}
\IEEEauthorblockA{\textit{Central Research and Technology} \\
\textit{Airbus}\\
Munich, Germany \\
svetoslav.duhovnikov@airbus.com} \\
\IEEEauthorblockN{Dominic Schupke}
\IEEEauthorblockA{\textit{Central Research and Technology} \\
\textit{Airbus}\\
Munich, Germany \\
dominic.schupke@airbus.com}
}

\maketitle
\copyrightnotice

\begin{abstract}

Cellular communication technologies such as 5G are deployed on a large scale around the world. Compared to other communication technologies such as WiFi, Bluetooth, or \gls{uwb}, the 5G communication standard describes support for a large variety of use cases, e.g., \gls{iot}, vehicular, industrial, and campus-wide communications. 
An organization can operate a Private 5G network to provide connectivity to devices in their manufacturing environment.
\gls{plkg} is a method to generate a symmetric secret on two nodes
despite the presence of a potential passive eavesdropper.
To the best of our knowledge, this work is one of the first to implement \gls{plkg} in a real Private 5G network.
Therefore, it highlights the possibility of integrating \gls{plkg} in the communication technology highly relevant for industrial applications.
This paper exemplifies the establishment of a long-term symmetric key between an aerial vehicle and IT infrastructure both located in a manufacturing environment and communicating via the radio interface of the Private 5G network.

\end{abstract}


\section{Introduction}

The promise of solving mathematically complex problems more efficiently has led to major developments in the field of quantum computing. A quantum computer reduces the time needed to solve some complex problems, which threatens existing systems relying on traditional symmetric and asymmetric cryptography algorithms such as \gls{aes}, \gls{dh}, and \gls{rsa}. 

Asymmetric cryptography algorithms, which are based on mathematical problems such as discrete logarithm and prime factorization, are not solvable on a classical computer in polynomial time but can be solved on a quantum computer using Shor's algorithm \cite{b1}. Increasing the key length used for symmetric cryptography algorithms such as \gls{aes} from 128 bits to 256 bits is one solution to safeguard against attacks from a quantum computer \cite{b2}. \gls{qkd} and \gls{pqc} are popular solutions to overcome threats posed by quantum computers.

Ranging from sensors to aerial vehicles, devices intended for mobile use cases rely on wireless communication for connectivity with varying levels of criticality and \gls{qos} requirements. Out of the wide variety of options available for establishing wireless connectivity, cellular technology in comparison is one of the most widely adopted with large-scale deployments around the world. This makes it a suitable candidate for many applications such as \gls{iot}, vehicular networks, or industrial networks. 

Due to the broadcast nature of wireless channels, securing communication between wireless nodes is important. \gls{pls}, as explored in this study, is a solution which could be integrated into many use cases. 
\gls{plkg}, a subset of \gls{pls}, generates a secret bit stream on two wireless nodes even in the presence of an eavesdropper. This method of generating bits can be implemented with low overhead as part of the channel estimation process typically carried out using the pilot signals. 
Recent developments in the area of cellular technology, \gls{evtol} aircraft, and quantum computing motivate this work; it explores a different approach to establish a secure link between communicating nodes against quantum threats.

\emph{Contribution: }This work aims to establish a secure communication link between an \gls{evtol} outside the manufacturing environment (e.g. when handed over to the customer) and the IT infrastructure present in the manufacturing environment using a symmetric key pair, which is generated while the \gls{evtol} was being manufactured. 
During the final stages of manufacturing, while the \gls{evtol} is still in the trusted manufacturing environment, \gls{plkg} is used to generate a symmetric key pair between the \gls{evtol} and the IT infrastructure. The generated key is used to secure communications between the \gls{evtol} and IT infrastructure at a later stage, i.e., after the \gls{evtol} moves out of the manufacturing environment, where key generation is more cumbersome. Beyond this use case, \gls{plkg} finds application in many more cases in the mobility domain and other industrial domains, where secure wireless connectivity is required.  
Hereby, the contribution of the work is an implementation of \gls{plkg} in a real Private 5G network and an evaluation of the implementation to assess the feasibility of using such a method for generating long-term keys of sufficient length and entropy. 

\emph{Outline: }Section \rom{2} discusses the features of wireless channels which enable \gls{plkg}, the steps involved in \gls{plkg}, and the contributions of this work and previous works. Section \rom{3} explains the use case and measures taken to implement \gls{plkg} in a Private 5G network. Details of the implementation and results are presented in Section \rom{4}. Section \rom{5} concludes this paper. 

\section{Background}

\gls{plkg} in a wireless channel is possible due to its frequency selective fading nature in environments where there is motion around the wireless nodes or in situations where one or both the nodes are moving. The bits generated in such channels are similar on both nodes as frequency-selective fading is reciprocal in nature within a period of time, referred to as coherence time, if no interferers are present. Hence, the frequency-selective nature of the wireless channel, if estimated by two communicating nodes within the coherence time, will be highly correlated in a non-interfered situation. Important features of a wireless channel that enable \gls{plkg} are as follows:
\begin{enumerate}
    \item \textit{Frequency-selective fading}: For generating dissimilar bits using \gls{plkg}, the wireless channel should affect each frequency component of the signal differently, i.e., it has to be frequency-selective.
    In a flat fading channel, repeating bits would occur, and the entropy of the generated bit stream would be low.
    \item \textit{Time-varying nature}: The frequency-selective nature of a channel should vary in time so that every channel estimate results in a different set of bits. A frequency-selective wireless channel static in time would produce highly correlated bit streams from consecutive channel probes.
\end{enumerate}
An important factor to consider is the presence of an eavesdropper. When two nodes, Alice and Bob, generate secret bits using \gls{plkg}, a third node Eve, who is aware of the algorithm used by Alice and Bob, cannot generate the same bit stream if it is located at least half a wavelength away from both Alice and Bob. This is due to spatial de-correlation in wireless channels as discussed in \cite{b3}.

\subsection{Physical Layer Key Generation}

\begin{figure}[h]
    \centering
    \includegraphics[width=0.45\textwidth]{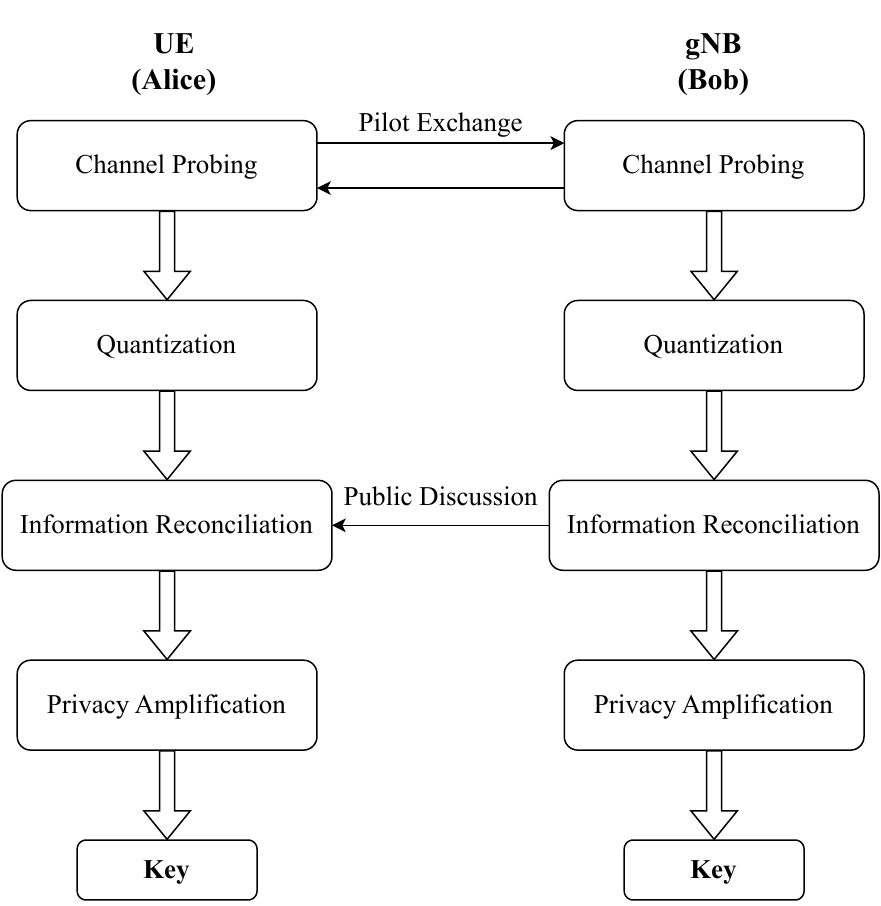}
    \caption{Sequence of steps involved in Physical Layer Key Generation.}
    \label{fig:PLKG1}
\end{figure}

To generate a symmetric secret bit stream on two communicating wireless nodes, steps illustrated in Fig.~\ref{fig:PLKG1} have to be performed by both nodes (Alice and Bob) described as follows:
\begin{enumerate}
    \item \textit{Channel Probing}: Alice and Bob exchange a signal with each other within the coherence time and capture the fading it undergoes due to the wireless channel, i.e., \gls{csi} estimation. Two ways to implement this step in practice are to measure (i) the Channel Frequency Response \cite{b8} and (ii) the Received Signal Strength \cite{b9}. The former is better regarding the amount of information extracted from the channel, and the latter is relatively easier to retrieve, e.g., from commercial WiFi \gls{nic} \cite{b10}.
    \item \textit{Quantization}: To convert the \gls{csi} estimates into bits, a quantization algorithm is used. It can be a single-level crossing-based or a multi-level quantization where the entire range of possible values of the \gls{csi} is divided into multiple regions, each corresponding to a predefined bit or sequence of bits.
    \item \textit{Information Reconciliation}: A mismatch in \gls{csi} estimate at Alice and Bob is expected due to asymmetric hardware defects and non-simultaneous measurements. This leads to mismatches in the bit stream generated by Alice and Bob. In \gls{plkg} and \gls{qkd} where the goal is to establish a symmetric key pair, an information reconciliation step is performed to correct mismatches or errors in the bit stream. \cite{b10} summarizes information reconciliation for \gls{plkg} in the existing literature.
    \item \textit{Privacy Amplification}: Due to the information reconciliation step, which often includes a public discussion, knowledge of the generated bit stream is leaked. A final step strengthens the generated bit stream to compensate for this and generate a cryptographic key.
\end{enumerate}
In this study, the channel frequency response is measured and used as \gls{csi}. The measured \gls{csi} is quantized to generate the bit stream. Information Reconciliation and Privacy Amplification are out-of-scope for this work. We focus on the Channel Probing and Quantization steps depicted in Fig.~\ref{fig:PLKG1}, as they depend on the nature of the wireless channel and features of the Private 5G network. Information reconciliation is circumvented by choosing suitable parameters for the quantization algorithm, i.e., larger sampling intervals in quantization are used to reduce bit mismatch at 5G Base Station (gNB) and \gls{ue}. For Privacy Amplification, we use a hash function to compress the quantized bits.

It should be noted that Private 5G enables a trusted environment for key generation (here the manufacturing site), since the spectrum is licensed locally for exclusive use. The access to this local spectrum as well as interferers can be managed by the spectrum owner. Therefore, reciprocity can be ensured.

Practical implementations of \gls{plkg} have been performed for WiFi, Bluetooth and \gls{uwb} \cite{b4, b5, b6, b7, b14}. Compared with these communication technologies, 5G by design supports a wider range of applications due to its advantages including larger coverage area and increased bandwidth and capacity.
Given the flexible nature of a 5G New Radio (5G-NR) physical layer, which includes multiple options for subcarrier spacing, density of reference signals in time and frequency, and flexible time slot configuration for uplink and downlink, a \gls{plkg} implementation in a Private 5G can be refined according to the environment. A simulation study for \gls{pls} implementation in \gls{lte} wireless standard has been presented in \cite{b12} and a practical implementation which measures entropy of generated bit stream by recording \gls{csi} on a single \gls{lte} node is presented in \cite{b13}.  Lack of practical 5G based PLKG testbench is discussed in \cite{b15}. This work demonstrates \gls{plkg} in a real Private 5G network by recording \gls{csi} on both nodes (gNB and \gls{ue}) and evaluates its feasibility in a manufacturing environment.

\begin{figure*}[h]
    \centering
    \includegraphics[width=0.95\textwidth]{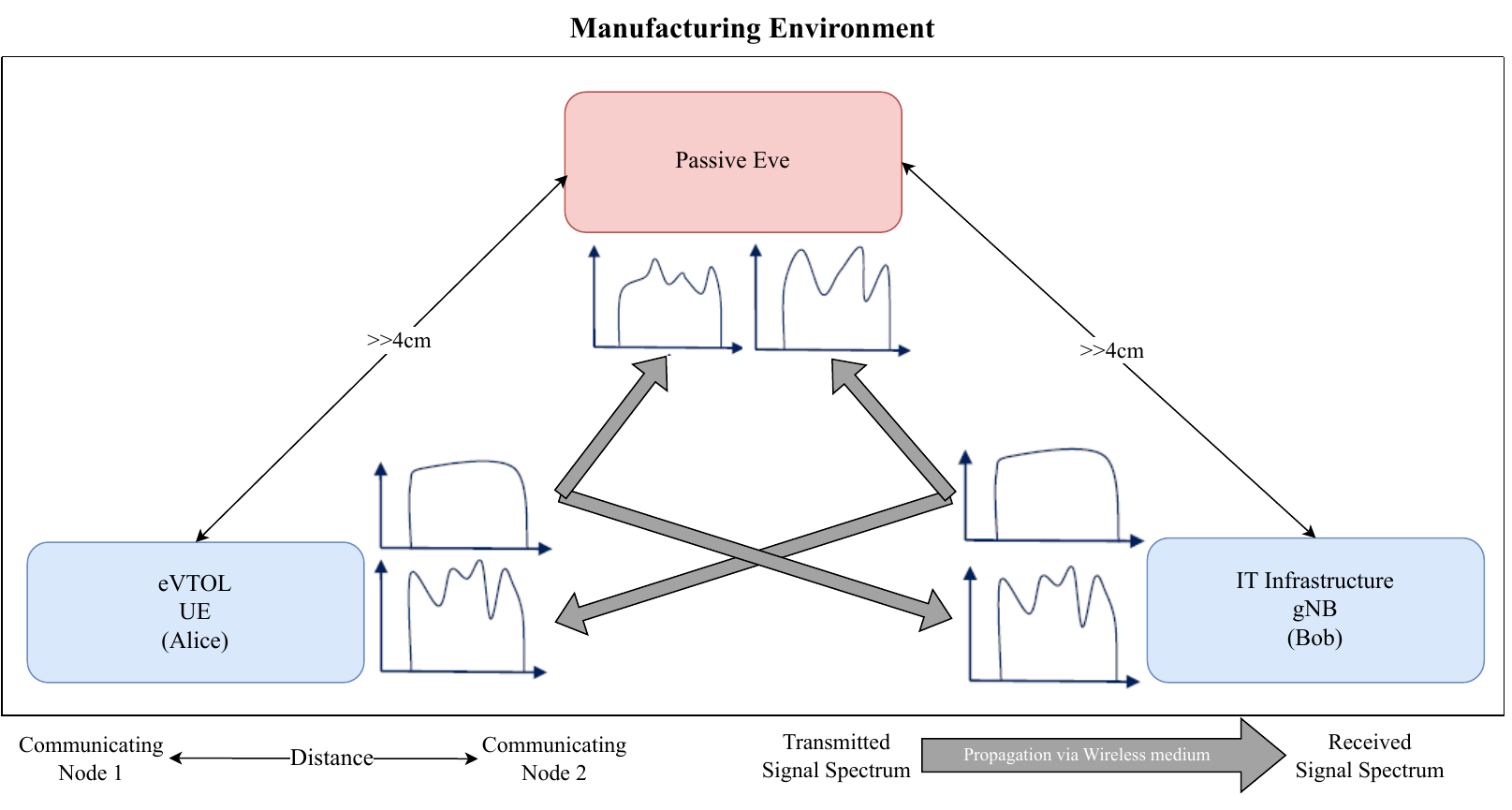}
    \vspace*{-4mm}
    \caption{An illustration of the use case where an \gls{evtol} (Alice) and IT infrastructure (Bob) both present in the manufacturing environment communicate via 5G-NR interface to generate a long-term symmetric key pair using \gls{plkg} in the presence of a passive eavesdropper.}
    \label{fig:UseCase}
\end{figure*}

\section{Concept}
\gls{plkg} can be used in mobile ad-hoc networks, wireless sensor networks, and wireless local area networks, where the devices can use generated keys immediately. This section explains the usage of the generated keys via PLKG as long-term keys used to establish a secure link between an \gls{evtol} and IT infrastructure in the manufacturing environment. This work aims to explore the applicability of PLKG in the presented use case. 
Parameters considered in this work within a Private 5G network to enable \gls{plkg} are explained in the second part of this section.

\subsection{Use Case}
\gls{plkg} is used to generate a symmetric key pair on an \gls{evtol} and IT infrastructure of the manufacturing environment. The manufacturing environment is a final assembly line where hardware and software components are put together to assemble the \gls{evtol}, after which it is delivered to the customer. During this assembly process, \gls{plkg} is carried out to generate a key pair for the \gls{evtol} and the IT infrastructure of the manufacturing environment. The environment is considered to be dynamic, with robots and personnel moving around the \gls{evtol} during \gls{plkg}. The manufacturing environment has a Private 5G network operating on the 5G-NR n78 band in \gls{tdd} mode. A benefit of using \gls{tdd} mode is that it enables both gNB and \gls{ue} to probe the same set of frequency components separated in time.
The generated keys are used by the vehicle to establish a secure communication link with the manufacturing environment for operations such as software updates and data offloading during its regular maintenance, which is not necessarily in an environment controlled by the manufacturer.
The advantages of using \gls{plkg} in this scenario are as follows:
\begin{enumerate}
    \item \textit{Information Theoretically Secure}: secret bits generated via \gls{plkg} are information-theoretically secure, i.e., they are secure against threats posed by quantum computers.
    \item \textit{Low Overhead Algorithm}: Sufficiently long bits can be generated via \gls{plkg} with low overhead. \gls{csi} estimates can be extracted from data and reference signals which are transmitted for normal communication between nodes, i.e., there is no need for an exclusive session for \gls{plkg} to take place. 
    \item \textit{Automation}: The entire process of \gls{plkg} can be fully automated. Hence, no involvement of personnel is needed and the time during which a given \gls{evtol} carries out \gls{plkg} can be hidden or randomized as well. 
\end{enumerate}

Eve in the manufacturing environment is assumed to be situated at a location \(\gg 4cm\) (half a wavelength of carrier frequency in 5G n78 band is around 4cm) from both the \gls{evtol} and wireless terminal of the IT infrastructure as shown in Fig. \ref{fig:UseCase}. To obtain similar \gls{csi}, Alice and Bob exchange pilot signals within the coherence period of the wireless channel. A passive Eve located \(>> 4cm\) from Alice and Bob records a different \gls{csi} as depicted in Fig. \ref{fig:UseCase}. This results in an uncorrelated bit stream generated by Eve \cite{b3, b11}. 

\subsection{PLKG in 5G}

To implement \gls{plkg} in a Private 5G network, a setup with an Amarisoft Classic gNB, a Raspberry Pi with SIM8200EA-M2 5G HAT as \gls{ue}, and two USRP B210 \gls{sdr} as recording devices were built. The \gls{sdr} recorded raw IQ samples at the antenna port of both devices from which relevant symbols were extracted and demodulated.   
The gNB and \gls{ue} communicated via a 5G link on the n78 band in \gls{tdd} mode. A bandwidth of 20MHz was used with 30kHz subcarrier spacing, resulting in 612 subcarriers for communication. \gls{dmrs} were configured to have two occurrences within a slot with mapping Type A and length 2. Typically, for channel estimation, \gls{dmrs} symbols are used. In this study, \gls{dmrs} and \gls{qpsk} modulated data symbols were  used to get the channel estimate. Multiple QPSK modulated data symbols within a single frame were averaged to get a less noisy estimate of the channel as shown in Fig. \ref{fig:Reciprocal}. Time-sharing of the channel was configured such that the uplink and downlink took place for a duration of 2ms and 2.5ms contiguous blocks respectively in every 5ms time period.
The remaining 0.5ms was not used by uplink or downlink and occurred between an uplink and downlink block. Measurements were carried out in an indoor lab as well as in an open space outdoors. 

Ideally, to implement \gls{plkg} on 5G-NR enabled devices (\gls{ue} or gNB), the \gls{dmrs} symbols after demodulation can be used as an input to the quantization algorithm to generate the bit stream.

\section{Results}

Using the Private 5G network setup built for \gls{plkg}, experiments were performed indoors and outdoors to record \gls{csi} estimates in environments that in part mimic a manufacturing environment. The recorded \gls{csi} was quantized, and a bit stream was generated. Using a hash function, the bit stream was compressed and the respective estimate of bit level security was calculated based on Lempel–Ziv–Welch lossless compression algorithm.

\subsection{Quantization Algorithm}
A quantization algorithm generates bits from the \gls{csi} estimates. A two-level (L=2) quantization can be used where a threshold is defined, and bits are derived based on the \gls{csi} being above or below the threshold. This approach results in blocks of 1's and 0's when the channel does not change rapidly over consecutive subcarriers, i.e., consecutive frequency components. To overcome this challenge, the quantization region was divided into multiple levels on the y-axis denoted by L. In addition to multiple levels, the width of each interval was computed in two ways. One where all levels have the same width and another where all the levels are equiprobable as first proposed for \gls{plkg} in \cite{b5}. The number of levels L was varied from 2 to 16. The number of bits sampled from each \gls{csi} estimate is denoted by S which was varied from 2 to 16. Elimination criteria were also implemented to eliminate \gls{csi} estimates representing a static channel and in situations where received signals had a very low signal-to-noise ratio.

\subsection{Results}

A larger variance was observed in a dynamic channel, i.e., in an environment where people were moving around the gNB and \gls{ue}, as compared to a static channel where people and objects around the communicating nodes were stationary. Compared to a static channel, the variance in \gls{csi} estimates in a dynamic channel indoors and outdoors was 1.35 times and 3.0 times higher, respectively. In a manufacturing environment, it is assumed that moving people, objects, and robots create a dynamic channel.

Based on this, \gls{plkg} was carried out in a dynamic channel. After channel probing, \gls{csi} estimates were quantized and tested for bias, i.e., to check if an equal number of 1's and 0's were generated in the entire bit stream. Ideally, the bias should be close to 0.5. Bias in this study was observed to be closer to 0.5 when L increased beyond 2. This bias test is only used as an indicator to find an anomaly in the generated bit stream before the next test is applied.

To compute the upper bound of entropy for the generated bit stream, the Lempel–Ziv–Welch lossless compression algorithm was used. After compression, the size of the bit stream reduced to 0.2 to 0.1 times of the input bit stream, i.e., the bit stream generated after quantization. Table \ref{tab:results} summarizes the results of \gls{plkg} showing the number of bits generated before and after compression for equal width of quantization levels, L = 4, 7 and S = 3, 5, 7, 9.

\begin{table}[]
\tiny
\resizebox{0.4\textwidth}{!}{%
\begin{tabular}{|c|c|c|c|}
\hline
\textbf{L} &
  \textbf{S} &
  \textbf{\begin{tabular}[c]{@{}c@{}}Number of bits \\ generated after \\ quantization\end{tabular}} &
  \textbf{\begin{tabular}[c]{@{}c@{}}Number of bits \\ generated after \\ compression\end{tabular}} \\ \hline
4 & 3 & 1020 & 142 \\ \hline
4 & 5 & 1765 & 202 \\ \hline
4 & 7 & 2576 & 282 \\ \hline
4 & 9 & 3294 & 343 \\ \hline
7 & 3 & 1020 & 186 \\ \hline
7 & 5 & 1765 & 295 \\ \hline
7 & 7 & 2576 & 400 \\ \hline
7 & 9 & 3294 & 499 \\ \hline
\end{tabular}%
}
\centering
\vspace*{2mm}
\caption{Length of generated bit streams after quantization stage and after compression in a period of 5 seconds is shown. L denotes the number of levels used for quantization and S denotes the number of bits generated from each \gls{csi} estimate. }
\label{tab:results}
\end{table}

\begin{figure}[]
    \centering
    \includegraphics[width=\columnwidth]{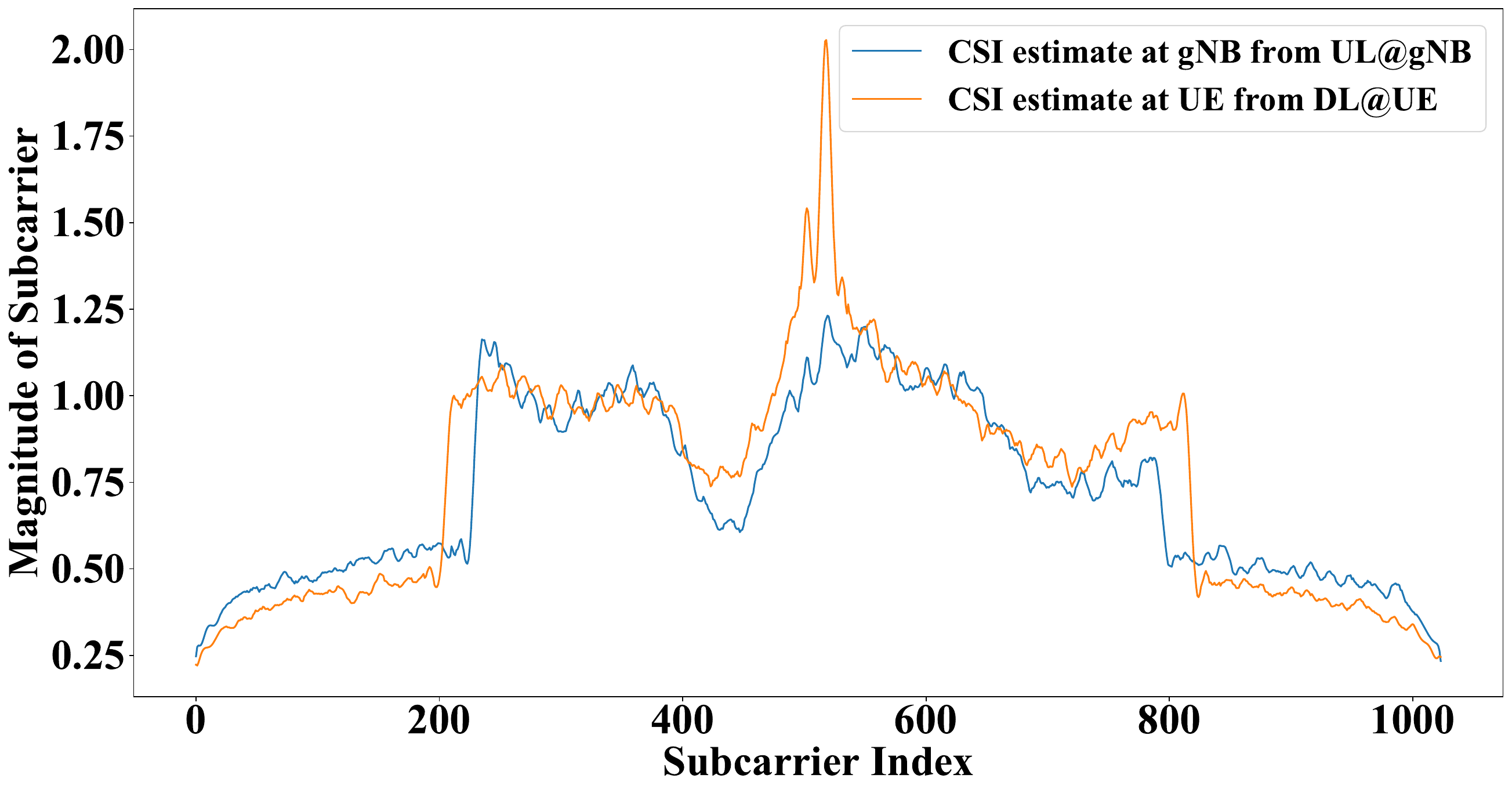}
    \vspace*{-7mm}
    \caption{\gls{csi} recorded using QPSK modulated data symbols at \gls{ue} (Alice) and gNB (Bob). The x-axis represents the subcarrier index. Subcarrier index 512 corresponds to the center frequency of transmission, i.e., 3.75GHz. The y-axis depicts the magnitude of the subcarrier.} 
    \label{fig:Reciprocal}
\end{figure}

For a symmetric key to be established between the nodes, the generated bit stream on both sides should be similar. 
Fig. \ref{fig:Reciprocal} shows the \gls{csi} estimates at gNB and \gls{ue} recorded within a duration of 10ms. For L = 4 and S = 3, similar bit streams were derived at gNB and \gls{ue}. For the considered quantization parameters at a channel probing rate of 10ms for a duration of 5 seconds, a maximum of 142 bits of entropy can be derived as seen in Table \ref{tab:results}.

\subsection{Discussion}
\textit{Limitations due to reciprocity:} The limit on L and S was due to the limitations with respect to reciprocity. For L = 7, the resulting bit stream at gNB and \gls{ue} had mismatches. To reduce the number of mismatches to 0, L = 4 and S = 3 were chosen. Other than increasing the duration of \gls{plkg}, for achieving a higher key generation rate, reciprocity between measurements at gNB and \gls{ue} must be improved by compensating for hardware defects. The poor reciprocal behavior can be in part attributed to the measurement setup which consisted of a \gls{sdr} tapping out signal from the antenna port, and to the use of a demodulator at a very early stage of development to extract CSI estimates. Impairments specific to this such as the DC offset in CSI estimate at \gls{ue} as seen in Fig. \ref{fig:Reciprocal} affect reciprocity. To overcome this, the DC offset was suppressed by interpolating to the nearest neighbours as seen for CSI estimate at gNB in Fig. \ref{fig:Reciprocal}. In this study, the correlation coefficient between \gls{csi} measurements at gNB and \gls{ue} varied from 0.1 to 0.85, the cause of the low correlation must be further investigated along with better methods to extract less noisy CSI estimates from the received signal. 

\textit{Eve:} Including an eavesdropper in future studies will help narrow down suitable quantization parameters. For example, in scenarios where \gls{csi} estimates of gNB and \gls{ue} have low correlation coefficients, reducing L and S will result in a similar bit stream. The extent up to which L and S can be lowered in the presence of an eavesdropper must be studied. Very low values of L and S will result in a similar bit stream at Alice, Bob, and Eve.

\textit{\gls{plkg} in Private 5G:} In a Private 5G network, as compared to a Public 5G network, physical layer parameters of 5G signals can be tuned for better \gls{plkg} performance. For a specific wireless channel, physical layer parameters such as subcarrier spacing, bandwidth, and \gls{tdd} slot configuration can be fine-tuned such that it can capture the frequency selective fading profile of the channel more effectively while maintaining reciprocity.

\section{Conclusion}

This work is one of the first to demonstrate a practical implementation of \gls{plkg} in a real Private 5G network. The feasibility of deriving a symmetric bit stream with an entropy of 142 bits on gNB and \gls{ue} in a duration of 5 seconds was shown. The chosen indoor and outdoor environments were found to have sufficient entropy to generate a 256-bit long key within a few seconds. A distinction between static and dynamic channel was made and it was found that dynamic channel was suitable for \gls{plkg} in both outdoor and indoor environment.
The result of this study motivates further development of 5G based PLKG testbench and investigation into \gls{plkg} for next generation cellular networks due to its wide range of applications.

\section*{Acknowledgment}
This work was partly funded by the Bavarian Ministry of
Economic Affairs, Regional Development and Energy as part
of the project 6G Future Lab Bavaria.

\end{document}